\documentclass[12pt]{article}
\usepackage{epsfig}
\usepackage{graphicx}
\usepackage{amsfonts}
\usepackage{latexsym}
\usepackage{amsmath,amssymb}
\usepackage{verbatim}
\usepackage[dvipdfm,hypertex]{hyperref}
\numberwithin{equation}{section}

\def\p{\partial}

\def\del{\nabla}

\def \Q{{\cal Q}}
\def\ts{\tilde \sigma}
\newcommand{\bea}{\begin{eqnarray}}
\newcommand{\eea}{\end{eqnarray}}
\newcommand{\be}{\begin{equation}}
\newcommand{\ee}{\end{equation}}
\newcommand{\ba}{\begin{align}}
\newcommand{\ea}{\end{align}}
\newcommand{\dr}{\Delta}
\newcommand{\non}{\nonumber}

\makeatletter
\let\over=\@@over \let\overwithdelims=\@@overwithdelims
\let\atop=\@@atop \let\atopwithdelims=\@@atopwithdelims
\let\above=\@@above \let\abovewithdelims=\@@abovewithdelims
\makeatother
\begin{document}
\begin{titlepage}
\today
\vskip 4 cm
\begin{center}

{}

{\Large \bf THE KERR-FERMI SEA}

\vskip 1.5cm {Thomas Hartman, Wei
Song, and Andrew Strominger}
\vskip 0.9 cm

{\em
Center for the Fundamental Laws of Nature\\
Harvard University, Cambridge, MA 02138, USA}\\

\end{center}

\vskip 0.6cm

\begin{abstract}
The presence of a massive scalar field near a Kerr black hole is known to produce
instabilities associated with bound superradiant modes. In this paper we show that for massive fermions, rather than inducing an instability, the bound superradiant modes condense and form a Fermi sea which extends well outside the ergosphere.  The shape of this Fermi sea in phase space and various other properties are analytically computed in the semiclassical WKB approximation.  The low energy effective theory near the black hole is described by ripples in the Fermi surface. Expressions are derived for their dispersion relation and the effective force on particles which venture into the sea.

\noindent

\end{abstract}

\vspace{3.0cm}

\end{titlepage}
\pagestyle{plain}
\setcounter{page}{1}
\newcounter{bean}
\baselineskip18pt


\tableofcontents

\section{Introduction}

  A massive scalar field has a dramatic effect on a rotating  black hole.
Small fluctuations of the scalar can grow exponentially - the so-called ``black hole bomb" \cite{bomb} - and classically destabilize the Kerr solution.  Recently it has been argued \cite {Arvanitaki:2009fg} that this phenomenon
both constrains the possible spectrum of light axions in our universe and predicts certain gaps in the mass-angular momentum spectrum of astronomical black holes.

The unstable modes arise from bound states with energies below the superradiant bound.
They are localized in a ``pocket" in the effective radial potential which is separated by a barrier both from infinity and the horizon.
One way to view the instability (in the particle picture) is this: thermodynamically the  superradiant modes want to form a Bose condensate in the pocket. This condensate then stimulates emission from the black hole, which leads to a larger condensate and even more emission. This runaway behavior is the black hole bomb.

It is not known if there are light scalars in the real world that could lead to such instabilities.
We do know however that there are light fermions. In this paper we investigate the effects of a light fermion on a rotating black hole in the semiclassical WKB limit, which turn out to be equally dramatic but qualitatively different. The WKB limit is relevant for fermions whose Compton wavelength is small compared to the black hole radius. For a neutrino near a solar mass black hole the WKB expansion parameter is around $10^{-7}$.

The leading WKB wave equation for a massive scalar and a massive fermion are identical
and equivalent to the massive geodesic equation. Therefore the fermions also see a pocket outside the black hole with bound  superradiant modes. However fermions cannot form a Bose condensate. Instead, in the standard Unruh vacuum, all the bound
 superradiant modes are filled, forming a Fermi sea which extends well outside the
ergosphere.  The low energy effective theory is described by ripples on the Fermi surface which can carry energies much below the fermion mass.
In the WKB approximation, many properties of this Fermi sea turn out to be analytically calculable.
We compute the number density of fermions, the fluid continuity equation and the
force on a test particle which couples to the sea fermions.

In principle, it is conceivable that our results are relevant for understanding the properties of astronomically observable black holes, and the Kerr-Fermi sea a component of dark matter. In this paper we consider only the simplest case of a free fermion on a fixed geometry, and a number of issues must be addressed before our results  can be applied to the real world.  Among these are the effects of fermion interactions, gravitational backreaction
and the rate at which the quantum state around a black hole formed by gravitational collapse approaches the Unruh vacuum.

In a completely different direction, our results may also be useful in the search for
holographic duals for rotating condensates for example of the kind created in the laboratory in \cite{Schweikhard:2004zz}.

Recently it has been argued that the  near-horizon deep-infrared region of extreme Kerr is dual to a two-dimensional conformal field theory (CFT) \cite{kcft}.  The Fermi sea described in this paper is localized a finite distance outside the horizon, and therefore is largely relevant at energies scales higher than those considered in \cite{kcft}. Nevertheless the sea does touch the horizon and so may be  relevant to  the  near-horizon CFT.   Fermi seas of this type  - localized outside the black hole - have been studied in the context of holographic superconductivity in \cite{Lee:2008xf,Liu:2009dm,liyu, sep}.

This paper is organized as follows. In section 2 we review the definition of the Unruh vacuum and describe the semiclassical limit.  In sections 3 and 4 we find the shape of the Fermi surface in phase space, compute the number density and other feature of the ground state, and derive the equation governing small fluctuations.  In section 5 we compute the force on a test particle which couples to the fermions along its worldline.   Although we anticipate that our results hold qualitatively for
general Kerr,  for the sake of brevity we restrict to the maximally rotating case. The non-extremal case is considered briefly in appendix A and details of the WKB approximation are provided in appendix B.

\section{Semiclassical limit of the Kerr-Unruh vacuum}

A Schwarzschild black hole has two well-known time-symmetric vacua: the Hartle-Hawking vacuum, which has a thermal gas of particles and a smooth stress tensor at the horizon, and the
Boulware vacuum which has no particles (according to static observers) and a singular stress tensor at the  horizon. In addition there is a third time-asymmetric but stationary vacuum,
the Unruh vacuum, which approximates the late-time asymptotic state around a  black hole formed by gravitational collapse.  It is characterized by no ``in" particles from past radial infinity  and a thermal flux of ``up" particles from the past horizon $H^-$. These initial conditions result in outgoing Hawking flux at future infinity and a smooth stress tensor at $H^+$. The Unruh vacuum describes a black hole emitting blackbody radiation into empty space.

For the Kerr black hole, analogs of the Hartle-Hawking and Boulware vacua are known generically $not$ to exist, see \cite{Unruh:1974bw, Leahy, Candelas, Frolov, Kay, Ottewill, Ottewill:2000yr} for discussion.
The basic problem is that one cannot
equilibrate modes with respect to the rotational potential outside the speed-of-light surface. The Unruh vacuum on the other hand does exist. For the sake of brevity we will specifically consider only the case of extreme Kerr except for appendix A where a few details of the general case are given. In the extreme case the Unruh vacuum is characterized by no incoming particles from past infinity and a ``thermal" flux at zero temperature with nonzero angular potential $\Omega_H={1 \over 2M}$ on $H^-$. This populates with unit probability all up
modes with energy below the superradiant bound: $E<m\Omega_H$ where $m=p_\phi$ is the angular momentum of the mode. More explicitly, in terms of the expansion of the
field operator for a Dirac fermion,
\be\label{unruhfield}
\Psi = \sum_{\ell,m}\int_{ E>0} d E\left( a_{ E\ell m}^{in}\psi^{in}_{ E m} +
    b_{ E \ell m}^{in\dagger} \psi^{in}_{- E-m\ell}\right)
   + \sum_{\ell,m}\int_{ E>m\Omega_H} d E\left( a_{ E\ell m}^{up}\psi^{up}_{ E m} +
    b_{ E \ell m}^{up\dagger} \psi^{up}_{- E-m\ell}\right) \ ,
\ee
where $\ell$ labels the angular eigenfunction, the Unruh vacuum is defined by
\be
a^{in}|0\rangle = b^{in}|0\rangle = a^{up}|0\rangle = b^{up}|0\rangle = 0 \
\ee
for all $a,b$.

The massive Dirac equation cannot in general be solved analytically for Kerr so, although we know in principle which modes are occupied in the Unruh vacuum,  it is difficult to determine quantities such as for example the stress tensor. In this paper we are interested in the leading semiclassical or WKB  approximation, in which case it turns out the leading behavior of many quantities of interest is analytically calculable. The semiclassical expansion parameter is the ratio ${ 1 \over \mu M}$ of the Compton wavelength of the particle to the Schwarzschild radius. For a neutrino near a solar mass black hole this is ${ 1 \over \mu M}\sim 10^{-7}$.

The semiclassical approximation begins with the phase space of a massive particle in the Kerr geometry which is labeled by the 6 coordinates $(r,\theta,\phi, p_r, p_\theta, p_\phi)$ . This phase space is constrained by the required existence of a real four vector $p_\nu$ obeying $p^\nu p_\nu=-\mu^2$, with $p_t=-E$,  and lying in the future light cone.\footnote{This constraint can be expressed \cite{MTW} as $g_{\phi\phi}E>-g_{t\phi} m $.  } We then define single particle states as cells of volume $\hbar^3$ in this phase space. A quantum state on the entire spacetime is specified by giving the occupation numbers of each of these states.\footnote{The notion of a particle depends on the choice of a coordinate system. In this paper we will use Boyer-Lindquist coordinates. Our operational definition of a particle in a given orbit is an occupied mode of the associated WKB wavefunction. This may become unphysical inside the ergosphere where static observers in Boyer-Lindquist coordinates are moving above lightspeed. In section \ref{s:sailing}, we will compute something more physical: the force on a probe which couples to the fermions comprising the Fermi sea.}   For fermions it can only be zero or one. For Dirac fermions, as shown in appendix B, we need four copies of this space  corresponding to spin up and spin down particles and anti-particles.

Cells in phase space are mapped to orbits by the action of the Hamiltonian, which in turn lift to pieces of WKB limits of solutions of the massive Dirac equation.
The cell is then filled if the corresponding semiclassical field mode is. Since the Unruh vacuum is stationary, all cells on the same orbit are filled with the same probability. Clearly a necessary condition for an orbit to be filled is
\be E<m\Omega_H.\ee

In general, the  WKB limit of a given solution has real ``under-the-barrier" regions as well as oscillatory regions.  More than one oscillatory region corresponds to more than one classical orbit separated by a potential barrier. Moreover, it may happen that one or more of the oscillatory
regions corresponds to an ``anti-orbit" with  momentum vector $p^\mu$ which lies in the past rather than the future light cone. Indeed this occurs in flat space. In the case of fermions these anti-orbits correspond to the negative-energy particles filling the Dirac sea.
In the case of Kerr we shall encounter anti-orbits with positive energy inside the ergosphere.

Orbits are characterized by the three conserved quantities $E$, $m$ and the Carter constant $\Q$ (defined in the next section) which are functions on phase space. Let us now consider the various possibilities in turn.

\subsection{$E>\mu$}
Orbits with $E>\mu$ have energy greater than their rest mass and are therefore not bound. They all reach infinity in the far past and/or future.
Those which do cross the horizon must have
\be \label{df} p_\mu\chi^\mu_H<0, \ee
where $\chi_H=\p_t+\Omega_H\p_\phi$ is the null generator of the horizon, as both $p$ and $\chi_H$ are future-directed at the horizon.  The condition (\ref{df}) may be rewritten as $E>m\Omega_H$, so we conclude that such  unbound orbits are unfilled.

If $E>\mu$ and $E<m \Omega_H$, the effective potential governing the WKB wavefunction has a barrier separating the near horizon region from the far region, under which the wave function must tunnel. In the semiclassical limit, this barrier is infinite, and the eigenmodes on the near and far side of the potential do not mix. Those on the far side
begin at spatial infinity and return there after reflecting off the barrier. Such modes are not populated in the Unruh vacuum.  Those on the near side begin and end at the horizon.
However if they have $E<m\Omega_H$ and cross the horizon, the momentum vector  must lie in the past light cone and they correspond to anti-orbits. These modes  are filled in the Unruh vacuum in the same sense that the negative energy modes in flat space are filled in the usual Dirac sea.

Away from the strict semiclassical limit, the WKB wave functions on the near and far regions have exponentially suppressed tunneling through the potential barrier. This is the well-known
Unruh-Starobinsky radiation of fermions \cite{Starobinsky:1973,Unruh:1974bw}.

\subsection{$E<m\Omega_H$ and $E<\mu$}
These orbits cannot reach infinity because their energy is less than their rest mass. They also cannot cross the horizon as they do not satisfy (\ref{df}). They are bound orbits which remain a finite distance outside the black hole. One expects that they are filled since they are near the
horizon which behaves like a reservoir for modes with $E<m\Omega_H$. We will see this is indeed the case.

\begin{figure}
\begin{centering}
\includegraphics{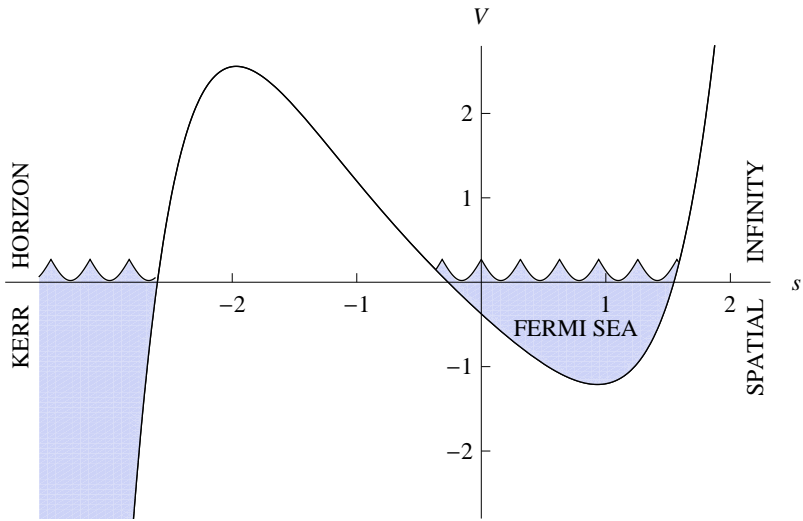}\\
\caption{Potential at $E={7\over8}\mu,\,m=2M\mu$ in units of $\mu^2 M^2$.  The horizon is at $s=-\infty$ and the boundary of the ergosphere at $s=0$.}\label{fig:potential}
\end{centering}
\end{figure}

To illustrate the point, let's take a specific orbit on the equator with say $E={7\mu\over8}$ and $m={\mu\over\Omega}$.
In the WKB limit the radial equation for the wave function becomes a Schr\"{o}dinger problem \be\label{se}(-\p_s^2+V)\psi=0,\ee with a potential \be V={\mu^2M^2}({15\over64}e^{2s}-{17\over16}e^s-{23\over64}+{7\over8}e^{-s}-{1\over16}e^{-2s}),\ee
where $s=\log(r/M-1)$, illustrated in figure \ref{fig:potential}.

This potential is of course the same as the one that appears in the geodesic equation for the corresponding orbits.
The horizon is at $r=M,~s=-\infty$. $V$ goes to $-\infty$ at the horizon and $+\infty$ far from the black hole.  In between there is a local minimum at $s_{min}=0.93$, and a maximum at $s_{max}=-1.97$. The zeros are at $(s_1,s_2,s_3)=(-2.61,-0.27,1.55)$.
The WKB solution  is
\be \psi(s)\sim  e^{-  \int^sds'\sqrt{V(s')}},\ee
where we take the branch of the square root which is positive at $s\to \infty$ where the wave function vanishes. At $s\to -\infty$ $\psi \sim e^{\pm i{\mu M\over 4} e^{-s}-i{7\mu \over 8}t +i2\mu M \phi }$.
Since the wave function emanates from $H^-$ rather than radial infinity  and has
$E<m\Omega_H$ the mode is filled in the Unruh vacuum.

The zeros of $V$  $(s_1,s_2, s_3)$ divide the real and oscillatory regions. The two oscillatory regions have $-\infty <s <s_1$ and $s_2<s<s_3$.    In the first region $g_{\phi\phi}E<-g_{t\phi} m $, so we have an anti-orbit. The filled mode does not correspond to a real particle in the sense that a geodesic observer very near the horizon will not see any particles.  Note that even though $E$, $m$ and $\Q$ are the same everywhere, it need not  be the case that the tangent vectors in both oscillatory regions are  both in the future light cone. The second region indeed has $g_{\phi\phi}E>-g_{t\phi} m $ and corresponds to a bound orbit in the equatorial plane (since we took $\Q=0$) around the black hole.  Since this orbit arises as the WKB limit of a filled mode, all the fermion states in this orbit are filled.

We note that the so-called Penrose orbits, which have $E<0$, can readily be shown \cite{MTW} to always have $E>m\Omega_H$. Hence filled orbits always have positive energy.

To summarize, the filled portion of the Fermi sea is the region of phase space
obeying
\be E<{\rm min}(\mu,m\Omega_H).\ee These are the bound
orbits with  energies below the superradiant bound. We will refer to the $E=m\Omega_H$
surface as inner and the $E=\mu$ surface as outer because the typical radii of orbits grows with angular momentum.

\section{The inner Fermi surface: $m\Omega_H<\mu$ }
The Fermi sea and surface each have two portions divided along the 5-dimensional hypersurface
$m=\mu/\Omega_H$ in phase space.  In this section we analyze the inner surface $m<\mu/\Omega_H$. The outer surface with
$m>\mu/\Omega_H$ is analyzed in section \ref{s:othersea}.
\subsection{The ground state}
The extreme Kerr metric is
\be\label{kerrmetric}
ds^2=-{\Delta \over \rho^2}\left(d t-M \sin^2\theta d\phi\right)^2+{\sin^2 \theta \over  \rho^2}
\left(( r^2+M^2)d \phi-M d t\right)^2+
{\rho^2 \over\Delta}d r^2+ \rho^2 d\theta^2
\ee
\be
\Delta\equiv (r-M)^2\:,\;\;\;\;\;
\rho^2\equiv r^2 +M^2\cos^2 \theta \ , \notag \ee where we take Planck units $G=\hbar=c=1$.
The horizon is at $r=M$. The angular velocity of the horizon is
\be \label{thm}
\Omega_H = {1 \over 2M}.
\ee
The 6-dimensional phase space for a particle in the Kerr geometry has coordinates  $(r,\theta,\phi,p_r,p_\theta,m)$ where we denote $m=p_\phi$.
The Fermi surface is a 5-dimensional hypersurface in the phase space, given by
\be \label{gh} E(r,\theta,\phi,p_r,p_\theta,m)=m \Omega_H\ee
where, for particle mass $\mu$,  $E=-p_t$ is the solution of the quadratic equation
\be \label{ff} p^\mu p_\mu=-\mu^2 \ee
such that $p^\mu$ is in the future light cone.
It is sometimes convenient to eliminate  $p_\theta$ in favor of the conserved quantity \cite{Carter:1968ks}
\be \Q=p_\theta^2+{(m-ME\sin^2\theta)^2\over \sin^2\theta}+M^2\mu^2\cos^2\theta-(m-ME)^2\ee
known as the Carter constant.  In terms of $\Q$ the general form of potential appearing in the radial equation $\Delta p_r^2+V=0$, or equivalently in the WKB Schr\"{o}dinger equation (\ref{se}) is \be
V=-{1\over\Delta}\left(E(r^2+M^2)-mM\right)^2+ \mu^2r^2+(m-ME)^2+\Q.
\ee
The insertion of the surface condition (\ref{gh}) into (\ref{ff}) yields
\bea \dr p_r^2&=&\chi m^2 -p_\theta^2-\mu^2\rho^2\\ &=&  {r^2+2Mr \over 4 M^2} m^2-\mu^2r^2 -\Q \notag \eea
where
\bea
\chi &=& {\rho^2\over \Delta \sin^2\theta}|\p_t + \Omega_H \p_\phi|^2\\
&=& {5\over 4} +{r\over 2M} + {r^2\over 4M^2} -  \csc^2\theta - {1\over 4}\sin^2\theta \notag
\eea
Therefore for the ground state the Fermi surface is at the real values of
\be \label{fs} p_r=\pm\sqrt{{\chi m^2-\mu^2\rho^2-p_\theta^2\over\dr}}\equiv \pm\sigma_0 \ee
where the $\pm$ denote the two branches of the hyperbola.  Note that near the north pole,
$\sin \theta \to 0$, and $\chi \to -\infty$ so $p_r$ can never be real. Hence the Fermi surface does not extend all the way to the poles.
\subsubsection{Number density}
Its interesting to compute the fermion density. The measure on phase space is, for $\hbar =1$,
\be { 1 \over (2\pi)^3} dr\wedge d\theta\wedge d\phi\wedge dp_r \wedge dp_\theta\wedge dp_\phi .\ee
The total number of fermions present is just the volume of the Fermi sea computed with this measure.  The density of fermions as a function of $(r,\theta,\phi)$ is
\bea \label{nd} N(r,\theta,\phi)&=&{1 \over  (2\pi)^3\sqrt{h}}\int_{\rm sea} dp_r \wedge dp_\theta\wedge dp_\phi \\  &=&{4 \over  (2\pi)^3\sqrt{h}}\int_{\mu \rho /\sqrt{\chi}}^{2M\mu} dm \int_0^{\sqrt{\chi m^2-\mu^2\rho^2}}dp_\theta  \sigma_0 \notag \\  &=&{ \mu^3 \over  (2\pi)^2\sqrt{h\Delta}}\bigl({4M^3\chi \over 3}-M\rho^2 +{\rho^3 \over 3 \sqrt{\chi}}\bigr)\notag \\ \notag\eea
where $h$ is the induced metric on a constant time slice and
\be \sqrt{h}= \rho \sin \theta \sqrt{(r^2+M^2)^2-M^2\Delta \sin^2 \theta \over \Delta} .\ee

For radii $r$ of order $M$, $\rho  \sim M$,  $\Delta\sim M^2 ,~~ \sqrt{h}\sim M^2$  and $\chi \sim 1$.
It then follows that $N \sim \mu^3$. This is just the statement that in the vicinity of the black hole there is of order one fermion per Compton cube. At large $r$, we have $\rho  \sim r $,  $\Delta\sim r^2 ,~~ \sqrt{h}\sim r ^2$  and $\chi \sim {r^2 \over 4 M^2}$. In this limit the range of the $m$ integral goes to zero and the leading and subleading terms in $N$ cancel. The leading nonvanishing term  gives $N\sim  \mu^3 ({ M \over r})^3 $. There is a log divergence at large $r$ in the total number of fermions.
\subsubsection{Angular momentum density}
It is easy to compute the angular momentum density $L$ by just inserting a factor of
$m $ in the Fermi sea integral (\ref{nd}).  One finds
\be L(r, \theta,\phi)= {\pi \mu^4 \over 4 (2\pi)^3 \chi \sqrt{h\Delta}}\bigl({4M^2\chi }-\rho^2\bigr)^2. \ee  Again there is a large $r$ cancellation, and the leading behavior is
$L\sim \mu^4 M ({ M \over r})^3 $. The angular momentum of the black hole itself is
$J=M^2$.  If we impose a large radius cutoff $r_c$, the ratio of the Fermi sea angular momentum to the back hole angular momentum is\footnote{The total angular momentum also has a logarithmic divergence at the horizon.   This divergence does not appear away from extremality, and is similar in form to well-studied divergences of the stress-tensor near the horizon of an extremal Reissner-Nordstrom black hole.  In the latter case the divergence appears to be rather benign and eliminated by gravitational backreaction - see \cite{trivedi} and references therein.  A similar outcome may pertain to extreme Kerr.}
\be {L \over J}\sim \mu^4 M^2 \ln {r_c\over M}.\ee

In Planck units, the mass of the sun is $10^{38}$, while a typical neutrino mass is  $10^{-31}$.
Hence for the neutrino sea around a solar mass black hole, for $L$ to be of order $J$ we need ${r_c \over M}\sim 10^{10^{48}}$. For $r$ of order $M$,  ${L \over J}\sim 10^{-48}$.
In this example the semiclassical expansion parameter ${1 \over \mu M}\sim 10^{-7}$ is very small.  The computation of the energy density is a bit more involved but one expects on dimensional  grounds $E \sim \mu^4 ({ M \over r})^3 $ which is also small. Hence the backreaction of the Fermi sea on the geometry is suppressed.
\subsubsection{$\langle \bar\Psi \Psi \rangle$}\label{s:psibarpsi}
In the WKB approximation (discussed in appendix B), the current for a single particle mode $\psi_\pm$ labeled by $p^\mu$ is
\be\label{jrel}
 \bar\psi \gamma^\mu \psi = \pm {1\over \mu} \bar\psi\psi p^\mu + O(\hbar)\ \ ,
\ee
where the sign is positive for particles and negative for antiparticles.  Because the sea has an equal number of particles and antiparticles, the total current vanishes
\be
\langle \bar \Psi \gamma^\mu \Psi  \rangle= 0 \ .
\ee
In the semiclassical approximation, $\langle \bar\Psi\Psi\rangle $ has  real and positive contributions from both particles and antiparticles. The expectation value
is given by summing (\ref{jrel}) over the Fermi sea,
\bea
\langle \bar\Psi\Psi \rangle &=& {4\mu\over (2\pi)^3\sqrt{-g}} \int_{sea} dm dp_\theta dp_r {1\over p^t}\\
&=& {8\pi \mu \rho^2\over (2\pi)^3\sqrt{-g\Delta}}\int_{\mu\rho/ \sqrt{\chi}}^{\mu/ \Omega_H}dm \left(m\Omega_H - E_{min}\right)\notag
\eea
where the factor of $4$ comes from particles and antiparticles of each spin, and
\be
E_{min} = {1\over \alpha}\left(2M^2 r m + \rho\sqrt{\Delta(\rho^2m^2/\sin^2\theta) + \Delta\alpha\mu^2}\right)
\ee
with
\be
\alpha \equiv -2M^2 r {g_{\phi\phi}\over g_{t\phi}} =  (r^2 + M^2)^2 - \Delta M^2 \sin^2\theta
\ee
is the minimum allowed energy of an orbit given $r,\theta,\phi,m$. Doing the final integral we find
\bea \label{pbp}
 \langle \bar\Psi\Psi \rangle&=& {2 \mu^3 \rho^2\over(2\pi)^2 \sqrt{-g\Delta}}\biggl[\left(\Omega_H + {g_{t\phi}\over g_{\phi\phi}}\right){m^2\over \mu^2} - {m\rho^2\sqrt{\Delta}\over\alpha \mu \sin\theta}\sqrt{{m^2\over \mu^2} + {\alpha \sin^2\theta\over \rho^2}} \\
& & \ \ \ - \sin\theta\sqrt\Delta\log\left({m\over\mu} + \sqrt{{m^2\over\mu^2} + {\alpha\sin^2\theta\over\rho^2}}\right)\biggl]\biggl|_{m=\mu\rho/\chi}^{m=\mu/\Omega_H} \notag
\eea

\subsection{Small fluctuations}

In this section we derive the first-order differential equation governing the propagation of small disturbances of the Fermi sea. Denoting the top of the sea by $\sigma = p_r(r,\theta,\phi,p_\phi,m)$, the general equation for the evolution of the Fermi sea is
\be \label{dft}
\p_t \sigma = [p_r, H] + \p_{p_\phi}\sigma [p_\phi,H]+\p_{p_\theta} \sigma [p_\theta,H]- \p_r \sigma [r,H] - \p_\phi \sigma [\phi,H] -\p_\theta \sigma [\theta,H]\\
\ee
where $p_\phi=m$ and $H=E$.   The second term vanishes. Now we expand
\be
\sigma = \pm \sigma_0 + \tilde{\sigma}^\pm.
\ee
(\ref{dft}) then becomes the fluid continuity equation
\be \label{fsb}\p_t\tilde{\sigma}^\pm\pm\p_r(v^r \tilde{\sigma}^\pm)+\p_\phi(v^\phi\tilde{\sigma}^\pm)+\p_\theta(v^\theta\tilde{\sigma}^\pm)=0\ee
which is related to conservation of $\bar\Psi \gamma^\mu \Psi$.
Here
\be
v^i = {\p H\over \p p_i}
\ee
\bea v^r&=&{\Delta\sqrt{r^2+2Mr-{4M^2\over m^2}(\mu^2r^2+\Q)}\over r(r^2+Mr+2M^2)+M^2(r-M)\cos^2\theta}\\ v^\phi&=&{2M(r-M\cos^2\theta)\csc^2\theta\over r(r^2+Mr+2M^2)+M^2(r-M)\cos^2\theta}\\
v^\theta&=&{2M(r-M)\sqrt{\Q-\cos^2\theta[\mu^2M^2+m^2({1\over\sin^2\theta}-{1\over4})]}\over m[r(r^2+Mr+2M^2)+M^2(r-M)\cos^2\theta]}\eea

\section{The outer Fermi surface: $m\Omega_H>\mu$}\label{s:othersea}
In section 3 we found the inner Fermi surface  for $m\Omega_H< \mu$ given by the condition $E = m \Omega_H$. Here we describe the case $m\Omega_H>\mu$ in which case the Fermi surface is at $E=\mu$ where the orbits become unbound. In phase space the equation $E = \mu$ becomes\bea p_r&=&\pm\sigma_0=\pm{1\over\Delta}\sqrt{c_2m^2-\mu c_1 m+\mu^2c_0-p_\theta^2\Delta}\\
c_2&=&M^2-\Delta\csc^2\theta\notag\\
c_1&=&4rM^2\notag\\
c_0&=&2Mr(r^2+M^2)\notag
\eea
Outside the ergosphere $c_2<0$ and the resulting contribution to the number density is
\bea\label{yy} \tilde N&=&{4\over (2\pi)^3\sqrt{h}}\int_{2M\mu}^{\mu p_+} dm \int_0^{\sqrt{(c_2m^2- \mu c_1m+\mu^2c_0)/\Delta}} dp_\theta \sigma_0\\&=&
{\pi\mu^3\over 2(2\pi)^3\sqrt{h}\Delta^{3\over2}}\left[2 c_0(p_+-2M) + c_1(4M^2 - p_+^2) - {2c_2\over 3}(8M^3 - p_+^3)\right]
\\p_+&=&{\sin\theta\left(-2M^2r\sin\theta+\sqrt{2Mr}(r-M)\rho\right)\over(r^2-2Mr+M^2\cos^2\theta)}\eea At radius of order $M$, the number density is also of order $\mu^3$ as in (\ref{nd}) so $N\sim \tilde N$. However at large radius (\ref{yy}) falls off more slowly than (\ref{nd}):
\bea
\tilde N &\sim& \mu^3({M\over r})^{3\over2}, \quad \hbox{as}\quad r\rightarrow\infty
\eea
The angular momentum density goes as $ L\sim \mu^4 M({M\over r})$, while the energy density goes as $ E\sim \mu^4 ({M\over r})^{3/2}$.
For $r\sim M$, the number, angular momentum and energy densities are comparable to those of the inner sea. However they fall off more slowly at large $r$.

The computation of $\langle\bar\Psi \Psi \rangle$ is identical to that of the inner Fermi sea, except that in the final answer (\ref{pbp}) the limits of integration are $m=\mu/\Omega_H$ to $m=\mu p_+$.

\section{Sailing on the Fermi sea}\label{s:sailing}
In this section we consider the effect of the Fermi sea on the trajectory of a particle with worldline coupling
\be \label{rf} \lambda \int d\tau \bar\Psi \Psi \ee
to first order in the coupling $\lambda$ which we take to be small. Here $\tau$ is the proper time along the particle trajectory $X^\mu(\tau)$, which implies $g_{\mu\nu}\p_\tau X^\mu \p_\tau X^\nu=-1$. There are two effects to consider: the change $\delta E_F$ in the energy of the Fermi sea and the change $\delta E_P$ in the energy of the test particle. Both of these result in an effective force on the test particle, and a modification of the geodesic equation if the particle is in free motion.  We will assume in this section that the particle motion is slow enough so that effective description of the Fermi sea in terms $\sigma$ given by (\ref{fsb}) and the semiclassical description are valid.

The change in the particle energy is read directly off of the interaction (\ref{rf}). It is nonvanishing at leading order in $\lambda$ because $\bar \Psi \Psi$ has a nontrivial vev, as computed for the inner sea in section 3.
One finds simply
\be \delta E_P=\lambda {d \tau \over dt}\langle \bar \Psi \Psi \rangle \ee

To compute $\delta E_F$, note that the presence of the particle modifies the Dirac equation to
\be \gamma^\mu\nabla_\mu \Psi-(\mu+\delta \mu)\Psi=0 \ee
with
\be \delta \mu={\lambda \over \sqrt{-g}} \int d \tau \delta^4(x^\mu-X^\mu(\tau)={\lambda \over \sqrt{-g}}{d\tau \over dt}\delta^3(x^i-X^i(t))\ee This in turn modifies the equation for the Fermi surface, which at linear level can be
expressed as a source term for $\ts$.  The modification is found simply by replacing $\mu$
by $\mu +\delta \mu$ in (\ref{fs}). This gives for the inner sea
\be \ts^\pm=\mp{\lambda \mu \rho^2  \over \sqrt{(-g)\Delta(\chi m^2-\mu^2\rho^2-p_\theta^2)}}{d\tau \over dt}\delta^3(x^i-X^i(t))\ee
The contribution to the energy of such a ``dimple"  from every $m$ and $p_\theta$ is
\be  -{\lambda m \mu \rho^2  \over M  \sqrt{(-g)\Delta(\chi m^2-\mu^2\rho^2-p_\theta^2)}}{d\tau \over dt}  \ee
Now we integrate over $p_\theta$ and $m$ and multiply by $4$ (for particles and antiparticles of each spin) to get the total energy.  This gives, where $\chi$ is positive,
\be \delta E_F= -{2 \lambda \mu^3 \rho^2(r-2M\cot^2 \theta) \over (2\pi)^2\chi  \sqrt{(-g)\Delta}}{d\tau \over dt} \ee
The gradient of this gives a force which vanishes at infinity and diverges on the horizon.
At large $r$, $\delta E_F$ goes as ${\mu^3 \over r^2}$  and the force as ${\mu^3 \over r^3}$. The sign depends on the sign of $\lambda$.

There is also a dimple in the outer sea.
Outside the ergosphere the energy is \bea \delta \tilde E_F =-{4\lambda\mu^3\rho^2\left(\sin\theta\sqrt{2Mr}\rho-2M(r-M\cos^2\theta)\right)\over(2\pi)^2\sqrt{-g\Delta}(r^2-2Mr+M^2\cos^2\theta)}{d\tau\over dt}\eea
At large $r$, $\delta \tilde E_F$ goes as $\mu^3r^{-{3\over 2}}$, and the force goes as $\mu^3 r^{-{5\over2}}$.

We note that the coupling (\ref{rf}) has the special property that, at first order, the presence of a particle only produces a dimple in the Fermi surface which follows the particle trajectory, but not a propagating wave.  For more general couplings -- such as $\int \bar \Psi \gamma_\mu\Psi d X^\mu$ -- we do expect waves to be produced and propagate according to the fluid continuity equation (\ref{fsb}).

\section*{Acknowledgements}
We are grateful to Sean Hartnoll, Gary Horowitz, Alex Maloney, John McGreevy, and Eva Silverstein for useful conservations.  This work was supported in part by
DOE grant DE-FG02-91ER40654 and the Harvard Society of Fellows.

\appendix

\section{Non-extremal Kerr}
We now consider a black hole at temperature $T_H$, with rotational parameter $a = J/M$, horizon at $r=r_+$, and
\be
\Delta \equiv r^2 + a^2 - 2 M r \ , \quad \rho^2\equiv r^2 + a^2\cos^2\theta \ , \quad \Omega_H = {a\over r_+^2+a^2} \ .
\ee
The inner Fermi surface, defined by setting $E = m \Omega_H$ in the geodesic equation, is
\be
p_r = \pm \sqrt{\chi m^2 - \mu^2 \rho^2 - p_\theta^2\over \Delta}
\ee
where
\bea
\chi &=& {\rho^2\over \Delta\sin^2\theta}|\p_t + \Omega_H \p_\phi |^2\\
&=& {[\Omega_H(r^2+a^2)-a]^2\over \Delta} - {(1-a\Omega_H \sin^2\theta)^2\over\sin^2\theta} \ .
\eea
At finite temperature the Fermi surface is smoothed out, but modes with
\be
0 < E < m\Omega_H  \ , \quad
{|E - m \Omega_H|} \ll T_H
\ee
are filled with unit probability.

\section{WKB limit of the Kerr-Dirac equation}
The WKB limit of the Dirac equation in a general curved spacetime is considered, for example, in \cite{Audretsch:1981wf,Rudiger:1981uu}.  Expand the fermion wavefunction as a series in $\hbar$ ,
\be \psi=\exp(-iA/\hbar)\sum_{n=0}^\infty(-i\hbar)^na_n(x) \ .\ee
Plugging this into the Dirac equation, we have the leading order equation
\be(\gamma^\mu A_{,\mu}-\mu)a_0=0 \ .\ee
The existence of a nontrivial solution for $a_0$ requires
\bea 
 A^{,\mu}A_{,\mu}&=&-\mu^2 \ .
\eea
Therefore $p_\mu=- A_{,\mu}$ is a congruence of timelike geodesics \cite{Audretsch:1981wf,Rudiger:1981uu}.  Focusing on a single worldline of the congruence, for modes moving forward in time we can adjust the local tetrad so that
\be p_a\equiv e_a^\mu p_\mu=(-\mu,0,0,0).\ee
Then the leading order positive energy solution is \footnote{Our gamma matrices are $\gamma^{\hat{i}}=\left(
                    \begin{array}{cc}
                      0 & \sigma^{\hat{i}} \\
                      -\sigma^{\hat{i}} & 0 \\
                    \end{array}
                  \right),\quad \gamma^{\hat{0}}=\left(
                    \begin{array}{cc}
                      1 & 0 \\
                      0 & -1 \\
                    \end{array}
                  \right)$, where $\sigma^{\hat{i}} $ are Pauli matrices.}
\be
\psi_+=e^{{i\over \hbar}\int dx^\mu p_\mu}\, \left(\alpha_1,\ \alpha_2,\ 0,\ 0\right)^T
\ee
along the worldline, where
\be D_\mu(p^\mu\alpha_{1,2}^2)=0\label{feq} \ .\ee
To leading order in $\hbar$, the fermion bilinears are
\bea
\bar\psi_+\psi_+ &=& |\alpha_1|^2+|\alpha_2|^2\label{ppls} \\
j^\mu_+ &=& \bar\psi_+\gamma^\mu\psi_+ = {\hbar\over 2 \mu}\left[(\del^\mu \bar\psi_+)\psi_+ - \bar\psi_+ \del^\mu\psi_+\right] + {\hbar\over 2 \mu}\del_\nu\left[\bar\psi_+\sigma^{\mu\nu}\psi_+\right] \\
&=& {p^\mu\over \mu}(|\alpha_1|^2
+|\alpha_2|^2)\label{jplsdn}
\eea
where we have used the Gordon decomposition to compute the current.

Similarly, the negative frequency solution is
\bea
&&\psi_-=e^{-{i\over\hbar}\int dx^\mu p_\mu}\, \left(0,0,\beta_1,\beta_2\right)\\
&&D_\mu(p^\mu\beta_{1,2}^2)=0
\eea
with the bilinears
\bea
\bar{\psi}_-\psi_-&=&-(|\beta_1|^2+|\beta_2|^2)\\
j^\mu_- &=& {p^\mu\over \mu}(|\beta_1|^2+|\beta_2|^2) \ .
\eea

So far, we have shown that to leading order the Dirac equation decouples into four components, particles and antiparticles of spin up and spin down, where each component travels along an identical particle geodesic.  We now specialize to the extreme Kerr metric.
There are three conserved quantities, the energy $E$, the angular momentum $m$, and Carter's constant $\mathcal{Q}$.
The Dirac field can be written as a mode sum
\be \Psi=\sum_{s=1,2}\int_{E > 0} dE dm d\mathcal{Q} \,(\tilde{a}\psi_++\tilde{b}^\dagger\psi_-)\ee
where the WKB wavefunctions $\psi_\pm$ and $a,\,b$ depend on $E,m,\mathcal{Q},s$. Comparing to (\ref{unruhfield}), we see that the Unruh vacuum has
\bea
\tilde{a}|0\rangle = \tilde{b}|0\rangle = 0  & & \quad \mbox{(in modes)}\\
\tilde{a}|0\rangle = \tilde{b}|0\rangle = 0 & & \quad \mbox{(up modes, $E >m \Omega_H$)}\notag\\
\tilde{a}^\dagger|0\rangle = \tilde{b}^\dagger|0\rangle = 0 & & \quad \mbox{(up modes, $E < m \Omega_H$)}\notag
\eea
Then we have
\bea
\non J^{\mu}&\equiv& \langle \bar{\Psi}\gamma^\mu\Psi\rangle \\&=&\int_{sea} dE dm d\mathcal{Q}\,\left(j_+^\mu - j_-^\mu \right)
\eea
\be\langle\bar{\Psi}\Psi\rangle=\int_{sea} dE dm d\mathcal{Q}\,{\mu\over p^t}\left(j_+^t + j_-^t \right) \ee

\end{document}